# The Mechanism for the Exceptionally High Tear Strength of Carbon Black / Hevea Natural Rubber Vulcanizates

G. Rong, J. Jiang, N. Schmitz, L. Jia and G.R. Hamed.

Department of Polymer Science

College of Polymer Science and Polymer Engineering

The University of Akron

Akron, OH 44325

## Abstract

Natural rubber vulcanizates containing 0-50 phr of a fine carbon black (N115, d ≈ 27 nm) were prepared and tensile strengths of normal (no pre-cut) and edge pre-cut specimens were determined. Normal tensile strengths of all vulcanizates were similar. At the relatively slow strain rate experienced wholesale by normal uncut specimens, all vulcanizates, prior to crack initiation, strain-crystallized sufficiently to be strong. However, pre-cut specimens experience increased strain rate at a cut tip. Magnification of the strain rate increases as cut depth **c** increases. Fracture in the gum NR and in vulcanizates with up to 14 phr of black occurred by simple forward crack growth from a cut tip, and all exhibited a critical cut size $\mathbf{c_{cr}}$, above which strength dropped abruptly. Furthermore, for these lightly filled samples, strength and $\mathbf{c_{cr}}$ ***decreased*** with increased black content. This indicates less strain-crystallization before rupture of pre-cut specimens when levels of black are low. This effect is attributed to rapid straining at a cut tip and hindering of the chain mobility necessary for crystallization. When black content was increased to 15 phr, with 1 mm < c < 2 mm, about 50% of specimens retained simple lateral fracture and were weak, but, the other 50% developed deviated cracks (knotty tearing) and were much stronger. With 50 phr of black, all pre-cut specimens exhibited knotty tearing and were significantly stronger than corresponding pre-cut gum specimens, especially at large **c**. High strengths with sufficient black levels are attributed to increased strain-crystallization and super-blunting (multiple cracks) at a cut tip. These inhibit forward crack growth. For carbon black to enhance strain-crystallization relative to the gum, it appears there must be enough of it to form a bound rubber/black network. If the black concentration is less than this percolation threshold, strain-crystallization is hindered at a cut tip.

## Introduction

The principle goal of this paper is to explain why sulfur-vulcanized, black-filled NR compounds are used in the belts of ***all*** tires (passenger car, truck, bus, airplane, off-road, etc.) produced by ***all*** manufacturers (Goodyear, Michelin, Bridgestone, etc.). Other parts of tires are based on a variety of elastomers and blends of these. The exclusive use of NR compounds in this critical, high-stress region of tires is due to their exceptionally high resistance to catastrophic tearing, now known to be related to rapid strain-crystallization of NR. (NR vulcanizates do not have a high resistance to ***low*** strain fatigue, where strain-crystallization does not occur.) In 1925, Katz[1], using X-ray diffraction, was the first to demonstrate that, although an undeformed NR vulcanizate was amorphous, it crystallized when sufficiently deformed.

The compound in the belt of a tire must resist tearing under extreme conditions involving high rates of local deformation (overload, low inflation, high speed, “pothole event”, etc.). Tires must not fail catastrophically; this is the reason NR compounds are always used in a radial tires’

"Achilles heel." (According to the National Highway Transportation Safety Administration (NHTSA) more than 95% of tire recalls involve belt edge separation.) Many tires are abused in service and develop substantial "separations." If such a tire resists catastrophic failure, its thumping noise and imbalanced feel alerts the driver that a replacement is needed. Otherwise, catastrophic "blow out" may ensue and cause the driver to lose control of the vehicle.

Studies of tear resistance at high local strain rate are needed. One simple way to increase local strain rate is to increase the length of an edge-cut in a tensile strip. A cut magnifies local strain rate (even though testing is carried out at the same "crosshead" speed), and varying its length provides a simple means of probing important rate effects. Results of such investigations are included in this paper. An example of the results is instructive. It will be shown that, without an intentional cut, the tensile strengths of gum (unfilled) and filled NR vulcanizates are similar. However, when a cut a few millimeters long is present, the filled rubber can be ***ten times*** as strong as the gum!

The starting point for understanding the fracture of solids is an analysis carried out by Griffith for the tensile fracture of an inorganic glass containing a crack of depth, **c**. Griffith[2] applied conservation of energy to fracture, assumed linear elasticity, and showed that

$$\sigma_b = \sqrt{\frac{2\gamma E}{\pi c}} \qquad (1)$$

where,

$\sigma_b$ = engineering tensile strength

E = modulus

$2\gamma$ = energy expended to break bonds that create surface per unit area of

the fracture plane (the fracture plane results in 2 fracture surfaces)

c = cut depth

The analysis assumes: (1). that no crack growth takes place until the stress obtains the breaking value $\boldsymbol{\sigma_b}$, at which time, a crack grows catastrophically forward. (2). the only strain energy expended during fracture is due to the bonds broken to create the fracture surfaces. (3). the entire input energy up to fracture is available to break bonds that result in fracture surfaces. (4) crack propagation occurs essentially perpendicular to the loading direction by growth of a single crack from a cut tip and (5). the shape of the crack tip is independent of crack length. Under these circumstances, $\boldsymbol{\gamma}$ determined using Equation 1, is like the surface energy determined from wetting experiments. This material is termed ***perfectly*** brittle.

Another perfectly brittle material is a highly crosslinked rubber (say, $M_c \approx 10^3$ g/mole). Such rubber is so fragile that a piece can be easily crumbled by hand! Why are highly crosslinked rubbers so very weak, even though more chemical bonds must be broken in order for them to be fractured? An analysis[3] like that used by Griffith has been employed to understand fracture of a tensile strip of rubber that contains an edge-cut:

$$\sigma_b = \sqrt{\frac{2GE}{kc}} \qquad (2)$$

where,

$\sigma_b$ = engineering tensile strength

E = modulus

k = “constant,” weakly dependent on breaking strain[4]

c = cut depth

G = the sum of the energy expended, $\mathbf{G_s}$ per unit area of fracture plane to break the bonds that create **s**urface area plus the energy expended in **b**ulk, $\mathbf{G_b}$; **G** is the total fracture energy per unit area of fracture plane

Under rapid tearing, **G** >> $\mathbf{G_s}$, sometimes by two or more orders of magnitude. Andrews[5] proposed that:

$$G = G_s \Phi(R, T, \varepsilon), \tag{3}$$

where **Φ** is an energy loss function, whose value depends on crack-propagation rate, **R**, temperature, **T**, and the overall strain level, **ε**, of the testpiece. Implicit in Equation (3) is the assumption that $\mathbf{G_b}$ is directly proportional to $\mathbf{G_s}$. Values of **G** at high rates increase more than an order of magnitude upon addition of carbon black to NR. Indeed, the intrinsic strength of the backbone bonds of an elastomer are not greatly altered by the presence of a particulate inclusion such as carbon black. It follows then that the substantial increase in strength for carbon-black-reinforced vulcanizates (under typical conditions) occurs primarily because of enhanced energy dissipation (increase in **Φ**).

The role played by hysteresis in enhancing strength may be seen by considering a crack propagating into rubber. Rubber-rubber network chains in elements in front of a crack tip first ***extend*** as the crack advances. Chains also slide over black surfaces, thus providing further resistance to extension and causing energy dissipation. Next, the extended elements ***retract*** as they enter the unstrained region behind the crack and further energy dissipation takes place. Network chains tend to return the elements to their original size. But, some of the movement of the chains is not reversible; this results in some set. Indeed, the amount of energy dissipation in cyclic deformation correlates directly with set.

Dissipative processes such as chain slippage along the surface of carbon black are important in reinforcement. A very strong interaction, in which slippage is largely prohibited, is undesirable. Additives that substantially promote bonding of carbon black to rubber have been shown to reduce both hysteretic losses and tear strength[6]. On the other hand, it is expected that extremely weak interactions are undesirable, since, even though slippage would readily occur, little energy would be expended in the process. Based on the above arguments, there must be an optimum level of adhesion between rubber and filler for the best reinforcement, just as there is a certain chain-chain interaction (crosslink density) which will maximize strength. Indeed, as noted earlier, too strong an interaction among network chains (*i.e.*, high crosslink density) results in embrittlement for all gum and filled vulcanizates – irrespective of the crosslinking chemistry. In short, excessive chain-chain interactions or chain-filler interactions diminish the molecular motions necessary for energy dissipation, and hence decrease strength.

Although frictional energy dissipation at the filler-chain interface and among chains, toughens an elastomer, there should be a <u>***threshold***</u> for this dissipation for compositions used in tires, which are subjected to dynamic fatigue. In this circumstance, (global) energy losses at small strains are minimized (→ low rolling losses, low heat build-up), but at locations of stress concentrations and resultant high strain, dissipative mechanisms should <u>***become***</u> active to inhibit crack growth. Gum NR is a good example; it is elastic at low strains, but becomes dissipative locally at a stress raiser and resists crack growth. Of course, the limitation of using gum NR vulcanizates in tires is their low modulus. An ideal elastomer for dynamic applications would be one with low $\mathbf{T_g}$ that is both stiff (without carbon black) and undergoes rapid strain-crystallization.

Now, we introduce a real-world example. Car and truck manufacturers have required tire companies to supply tires with low rolling resistance (i.e., low energy dissipation) during service. This must be carefully done, since, as explained above, strength may be substantially reduced when rubber is formulated to have low energy dissipation. An important example showing an unacceptable energy dissipation/strength tradeoff is the recent recall of tires by a major tire company due to tearing-out of tread lugs. Lower rolling resistance was achieved at the expense of adequate tear strength. Treads must retain high tear strength when formulated for low rolling resistance. For NR/black vulcanizates it is important that strain-crystallization be maintained especially under conditions of high severity. The rapidity of strain-crystallization and the melt temperature of strain-crystallites should be high. This depends on the level and type of crosslinking as well as backbone modifications during crosslinking and the type and amount of carbon black.

Finally, we consider the effect of rigid inclusions on the strength of rubber.

Rigid Inclusions (d = 1 μm)

The stresses around a rigid spherical inclusion; well bonded to an elastic ***continuum*** under tensile deformation were calculated by Goodier[7] in 1933. He showed that the maximum stress was near the two poles of the inclusion (in the deformation direction). This stress was triaxial, with all three stress components being about twice the tensile stress far from the sphere. Thus, a rigid inclusion magnifies stress around it. How, then, can many rigid spheres result in substantial strengthening? They cannot! Goodier's analysis is ***not*** applicable to black-filled, vulcanized rubber, because the size of carbon black nodules is comparable to the distance between crosslinks, thus, a continuum approach is not valid. It is impossible, no matter the inclusion/rubber adhesion, to obtain substantial strengthening (say, $\boldsymbol{\sigma_b}$ increase by 10x) of SBR by incorporating rigid inclusions that are 1 μm in diameter. However, if enough well-bonded, micron-sized rigid spheres are incorporated into rubber, $\boldsymbol{\sigma_b}$ can be increased modestly as shown below.

Because of complex interactions, it is impossible to calculate the stress-field in a continuum that contains a significant volume fraction (say ≈ 25 volume percent) of rigid micron-sized spheres. However, modulus has been successfully predicted for well-bonded, micron-sized spheres using the Guth-Gold equation[8]:

$$E_f = E_o(1 + 2.5v + 14.1v^2) \qquad (5)$$

where,

$E_o$ = modulus of the continuum

v = volume fraction of spheres

$E_f$ = modulus of the filled rubber containing the spheres

It is noteworthy that, in Equation 5, $\mathbf{E_f}$ does not depend on sphere diameter, **d**. Equation 5 is not valid for very small inclusions, where inclusion size and, hence spacing, are like the distance between crosslinks. When this is the case, $\mathbf{E_f}$ increases as **d** decreases, if filler particles are well bonded to the rubber. However, keeping **G** constant, if Equation 5 is used to calculate $\mathbf{E_f}$ and this value is substituted into Equation 2, the tensile strength $\boldsymbol{\sigma_{bf}}$ of a rubber that contains 25 volume percent micron-sized filler is calculated to be $\sigma_{bo}(E_f / E_o)^{1/2}$; this corresponds to a 60% increase in tensile strength. Results are shown in Figure 1, where the stress-strain curve for a gum SBR vulcanizate is compared to one that contains 25 volume percent of $CaCO_3$ ($d \approx 1$ μm). Also shown are stress-strain curves for black-filled SBR and NR, and for gum NR vulcanizates. These results have been previously discussed[9].

<u>Rigid Inclusions (d = 20 nm)</u>

Figure 2 is a diagram showing spheres on a square lattice. Equations are given to calculate various parameters. When d = 20 nm at 25 volume percent of spheres, the distance between particles is like the distance between crosslinks (dots in Figure 3); this negates the use of continuum mechanics. Interparticle spacing s = 5.6 nm, and all chains are within 3 nm of particle surfaces. Network chain mobility is reduced because of proximity to filler surface; crack opening is hindered. Although the filler distribution in this model is idealized, the basic principles hold for actual reinforced systems.

Particle spacing considerations can be used to explain the difficulty in dispersing very finely divided filler in high molecular weight elastomers. For d = 20 nm and v = 0.25, *s* = 5.6 nm (Figure 2). Uniform dispersion into an elastomer with M = 500K ($r_g \approx 13.4$ nm) would require a particle-to-particle spacing less than the elastomer coil size. Thus, for perfect dispersion the elastomer would need to distort, which is entropically unfavorable. Instead, poor dispersion results with regions of filler agglomerates separated by, at least, the elastomer coil size.

Dynamic mechanical analysis (DMA)[10] and solid-state NMR[11] have been used to probe changes in segmental mobility when carbon black is added to rubber. NMR indicates a strongly adsorbed layer of rubber; it is highly restricted. The peak in tangent delta (DMA) as a function of temperature shifts only slightly when carbon black is added to rubber. However, the breadth of the peak is widened (on the rubbery side) due to restrictions imparted by the black surface. Although $T_g$ (i.e., maximum in tangent delta) changes little with the addition of carbon black, the ***translational*** mobility of chains required to "open-up" a crack is hindered because of their proximity to particles.

Filler/Rubber Interactions

In addition to the requirement that filler particles must be sufficiently small to give substantial reinforcement, rubber/filler interactions are also important. It has been shown[12] that the interaction of polymers with surfaces may: increase, decrease or not change the glass transition temperature of the polymer (near the surface) compared to its bulk value. Of course, the segmental mobility of chains interacting with filler depends on the number and types of bonds formed between the filler and rubber. Many strong interactions will reduce mobility (increase $T_g$), while very weak ones will increase mobility (decrease $T_g$). Diene rubber chains interacting with tire-grade furnace blacks have reduced mobility--due to chain confinement (spatial restriction) and due to strong filler/rubber bonding (surface restriction). However, even fillers that exhibit weak interactions with rubber, can result in high tensile strengths, if the filler particles are small. Both graphitized[13] and fluorinated[14] blacks interact weakly with rubber, inferred from low levels of bound rubber; nonetheless, these fillers can impart high $\boldsymbol{\sigma_b}$ to SBR (The strengths based on the cross-section at break are higher for vulcanizates containing the modified blacks!), although in both cases, modulus and abrasion resistance are well below that obtained with unmodified black. Particles with diameters like the distance between crosslinks are ***necessary*** and ***sufficient*** to provide enough spatial restriction to give high tensile strength--at least at room temperature and normal strain rates. This statement is ***not*** valid for all mechanical properties. Indeed, when considering the effect of compound variations on the mechanical properties of rubber, it is imperative to recognize that strength and stiffness depend on the rate of deformation, temperature, environment, type of deformation and degree of deformation. Furthermore, in general, fracture properties depend on stiffness, since this may influence the stresses experienced by a rubber good in service. Unfortunately, this has not been well recognized in the literature, and, as a result, many false generalizations have been made.

It would be interesting to test the mechanical properties of vulcanizates that contain blacks with varying degrees of graphitization. There is expected to be some level of graphitization where "neutral" surface interactions would occur ("theta-surface"). This is defined as an interaction with the surface, where surface chains have the same mobility, and hence $T_g$, as those in the bulk. Much research is still needed to understand the role of spatial restriction and surface interactions on the physical properties of filled rubber.

In rubbers that have been highly reinforced by fine filler particles, matrix chains are expected to have multiple sites of interaction along the surface of the particles. Translation of chains requires the disruption of many chain/filler bonds, the number of which may be estimated. Assuming, conservatively, that one bond occupies 0.25 $nm^2$ of filler surface, and taking d = 20 nm and v = 0.25, it is calculated that there would be 6.6 ($10^{-3}$) moles of bonds per mL of matrix. The reasonable value of $M_c = 10^4$ g/mole for the matrix corresponds to 4.5 ($10^{-5}$) crosslinks per mL of matrix. Thus, ***the matrix has a considerably larger number of attachments to filler than it has crosslinks***. As a result, when breaking nano-composite rubber, more energy must be expended to disrupt chain-filler bonds than expended to disrupt rubber-rubber network bonds. Briefly, fracture energy is dominated by a "volume" effect, i.e., bulk energy dissipation. In principle, then there must be a maximum size of the "entity" that joins rubber chains and that results in substantial strengthening. If too small, the "entity" simply behaves like a crosslink. The smallest "entities" that have been shown to give substantial reinforcement are chemically attached ion-pairs, which form stiff domains about 3 nm in size in a hydrocarbon rubber[15]. In calculations carried-out using Figure 2, just 2% by volume of reinforcing entities (3 nm in size) will provide spacing of reinforcing entities like that found for entities with v = 0.20 and r = 15 nm.

## Experimental

Materials

1) Natural rubber: SMR CV60, Akrochem Corporation.
2) N115 carbon black: Cabot Corporation. The average primary particle size of N115 black is 27 nm. The DBP absorption value is 113 cc / 100 g.
3) Stearic acid: Harwick Chemical Company.
4) Zinc oxide: Akrochem.
5) Antiozonant PD-2: N-(1,3-dimethyl butyl)-N'-phenyl-p-phenylene diamine, Akrochem.
6) Antioxidant DQ: 2,2,4-trimethyl-1-2-hydroquinoline, Akrochem.
7) Akrowax™ Micro23: microcrystalline wax, Akrochem.
8) Santogard PVI: N-(Cyclohexylthio) phthalimide, Flexsys America.
9) Sulfur: Harwick Chemical Company.
10) TBBS: N-tert-butyl-1,2-benzothiazolesulfenamide, Flexsys America.

<u>Formulations</u>

Rubber compounds contained 0-50 phr of black. Each compound contained: 1.8 phr stearic acid, 3.5 phr zinc oxide, 1.5 phr Antiozonant PD-2, 1.0 phr Antioxidant DQ, 1.0 phr wax, 0.1 phr PVI, 1.75 phr sulfur, and 0.75 phr TBBS. Assuming densities of black and natural rubber are 1.8 g/cc and 0.9 g/cc, volume fractions **v** of black were determined (Table 1).

Mixing

Carbon black master batches first were prepared in a 250 mL Banbury internal mixer using a fill factor of 0.7. The rotor speed was 100 rpm. The mixing procedure is shown in Table 2.

Milling

The curatives (sulfur and TBBS) were added on a two-roll mill (Farrel, 15 cm diameter and 30 cm roll length). The milling procedure is given in Table 3. Compounds were milled for three minutes and then curatives were slowly added to the bank with alternating cuts. Sheets were taken off the mill after 10 end-roll passes at a 1.9 mm nip. Stocks were stored at room temperature at least 24 hours before vulcanization.

Vulcanization Characterization

Vulcanization kinetics were determined from rheometer curves using an Alpha Moving Die Rheometer (MDR) 2000 at 140°C and 3° arc. The cure time $\mathbf{t_c}$ **(100)** was the time to reach maximum torque.

Molding

Unvulcanized milled sheets (about 10 x 10 x 0.2 cm) were placed in the center of a window mold (160 x 160 x 0.5 mm). Then Mylar films and two aluminum plates were placed on each side. Sheets were cured at 140°C to $\mathbf{t_c}$ **(100)** in a Dake hydraulic press under a load of about 30 tons and then quenched in water.

Bound Rubber

Uncured black/NR mixtures were prepared using the same procedure as that used for cured vulcanizates, except that cure agents were not added during milling. About 1.0 g of uncured sheet (**$W_{initial}$**) was weighed and immersed in 80 ml of toluene for four days. The solvent was replaced twice during this extraction. After the extraction, the sol portion had been removed. The swollen gel was dried under vacuum at room temperature. Then, the dried sample was weighed (**$W_{dry}$**). Bound rubber content was calculated by the following equation:

$$Bound\ rubber\ (\%) = \frac{W_{dry} - \frac{phr\ of\ carbon\ black}{100 + phr\ of\ carbon\ black} \times W_{initial}}{\frac{100}{100 + phr\ of\ carbon\ black} \times W_{initial}}$$

Tensile Testing

ASTM D412 Type C dumbbells were cut in the milling direction from cured sheets. The width of the narrow section was 6.35 mm and the thickness was about 0.5-0.7 mm. Crosshead speed was 50 mm/min with an initial clamp separation of 65 mm. The nominal strain rate was 0.77 $min^{-1}$. At room temperature, strain was measured in the narrow section of a specimen by a mechanical “clip-on” type extensometer with an initial separation of 25 mm. At higher temperatures, strain was measured by the distance between two markers with an initial separation of 25 mm. For edge-cut specimens, the edge-cuts were introduced using a razor blade dipped into a soap solution to reduce friction. The cutting was done vertically to minimize difference of cut depth on each side. Cut depth was measured using a traveling optical microscope. Stress was calculated based on entire specimen width (6.35 mm).

Crack Pattern Photographs

After fracture, the two broken pieces were placed back together and photographs of crack patterns were taken with a Nikon D1X digital camera. An arrow in a picture indicates the tip of an initial razor cut. The designation **Ci** means crack "**i**" and **Ai** is the site of arrest of crack **"i"**.

## Results and Discussion

Tear Strength

Up until about 1922[16-19], systematic tear testing of rubber was not carried-out. Only hand tearing of a cut sheet was used for evaluation! Thus, simultaneous publication by four different persons[16-19] suggests an industry-wide recognition of the importance of tear testing in 1922-1923. Edge-cut tensile[18], T-peel[16, 19] and center slit tearing[17] were described.

Now, we show the great value of testing the tear strength of testpieces that contain edge cuts of various depth, **c**. Figure 4 shows the stress to break $\boldsymbol{\sigma_{bc}}$ at various **c** for a gum NR vulcanizate and for one that contains 50 phr (about 20 volume percent) of N115 black. Also shown, as the dotted lines, are their normal (no-cut) tensile strengths $\boldsymbol{\sigma_{bo}}$. Both rubbers have similar $\boldsymbol{\sigma_{bo}}$ ($\approx$ 26 MPa), but, when a cut had been introduced, great differences are seen. When **c** is ***just*** 0.1mm, the filled NR is now twice as strong as the gum. Up to c $\approx$ 0.7 mm, the filled NR has little change in $\boldsymbol{\sigma_{bc}}$, while the gum strength has dropped to about 6 MPa. At the largest c $\approx$ 3 mm, the filled rubber is almost 10x as strong as the gum! (The sudden drop in strength of the gum NR is discussed in a recent paper[20].) An important difference is seen in the way cracking took place from cut tips. In the gum testpieces, one crack initiated from a cut tip, then it suddenly propagated across a specimen. When precut black NR vulcanizates were strained, fracture was complex as shown in Figure 5 (Specimen #16, Figure 4). Upon straining, the cut tip grew slowly, very slightly forward, before it (A1) turned upward, grew rapidly and arrested. As further deformation occurred, no new cracks formed, until suddenly crack A2 ran in the opposite direction and stopped. Again, no further crack growth took place until rapid step growth of A3. These three cracks release strain energy at the crack front and delay the onset of the catastrophic crack C4. All three pre-cracks turn ***back*** toward the edge of the testpiece that contains the cut; this increases their stability. With cuts up to 0.7 mm in depth, multiple cracks cause the tear strength of pre-cut black-filled testpieces to be nearly equal to that of uncut specimens!

Crack splitting or turning occurs only when the region between particles is small enough. Widely spaced particles, inherent in a micro-composite rubber, result in simple growth around the "obstacle" through bulk-like matrix, but with nano-spaced particles "forward" cracking is inhibited and crack branching provides energy release by sideways **stable** cracking. This super-blunting mechanism reduces stress concentration and delays the onset of catastrophic crack growth.

Energy dissipation plays an essential role in "super-blunting." Hysteretic chain motions along filler surfaces result in high chain alignment locally between particles. This results in more uniform chain loading, so that critical strength anisotropy obtains and, hence, crack deviation takes place before the concerted chain breakage necessary for a crack to continue forward. Carbon black increases the tear strength of NR for two reasons: increased energy to break material in front of the crack front, plus the ability to knot. This is one reason tensile strength and tear strength do not correlate.

Crack growth into the direction of straining was first discovered in black-filled rubber by Pickles[21], who termed it "knotty tearing," because it reminded him of a crack that propagates around a knot in a tree. Similar behavior (i.e., knotty tearing) is seen in other tearing geometries. In trouser tearing, successive fast and slow crack growth (sideways knot formation) occur. In T-peel tearing, crack deviation (from simple forward growth) results in a fracture surface that is very rough. The ability of highly black-filled NR to form "knots" contributes greatly to its outstanding resistance to tearing. To the authors' knowledge, there is no rubber that resists tearing as well as sufficiently black-filled NR.

Figure 6 is a schematic that illustrates the mechanism of crack deflection (knotting) in black-filled NR. When there is enough chain alignment and crystalline fibril formation, such that the fracture energy for knotting is about 40% of that for forward growth, a crack is inhibited from growing forward and one grows sideways (knots) instead[22]. Strain-crystallinity alone does not provide enough anisotropy in strength to cause knotting; carbon black is needed. Indeed, amorphous, black-filled SBR vulcanizates can exhibit some knotty tearing, but not to the extent of black-filled NR. Gehman and Field[23] showed that enough carbon black promotes strain-crystallization of NR. They found that a typical gum vulcanizate of NR began crystallization at about 400% elongation, while one filled with 48 phr of channel black commenced crystallizing at 200% elongation. Thus, strain crystallization and carbon black are needed to obtain extensive knotty tearing. Returning to Figure 3, obviously, we want to know the behavior of $\boldsymbol{\sigma_{bc}}$ for compounds containing between 0 and 50 phr of N115 black.

Figure 7 compares $\boldsymbol{\sigma_{bc}}$ of the gum and a compound that contains 12 phr of black! It would be reasonable to expect that $\boldsymbol{\sigma_{bc}}$ with 12 phr of black would lie ***between*** the behaviors for 0 and 50 phr of black. But, the results are very different: (1). $\boldsymbol{\sigma_{bo}}$ is similar ($\approx$ 24 MPa) for the gum vulcanizate and for one with 12 phr of black. (2). When $c \geq 1.7$ mm, $\boldsymbol{\sigma_{bc}}$ for the two vulcanizates is also similar. (3). But, when $0.1 \text{ mm} < c < 1.7$ mm, the behaviors are markedly different! Testpieces with 12 phr of black are substantially ***weaker*** than the gum and these lightly filled samples exhibit a smaller critical value of cut depth $c_{cr} \approx 0.7$ mm. This means NR with 12 phr of black strain-crystallizes ***<u>less</u>*** than the gum! Similar strength, when c = 0 and c > 1.7 mm, is attributed to the long time before fracture that allows both rubbers to crystallize enough to be strong when c = 0 and, when c > 1.7mm, the strain rates at cut tips are so rapid that crack growth occurs before much crystallinity has developed. Furthermore, all specimens with 12 phr of black fail like the gum, i.e., by the forward growth of a single crack from cut tips.

The first demonstration that the strength of edge-cut NR specimens decreased, when black levels are low, was shown by Hamed and Al-Sheneper[24]. The authors are aware of no other tearing geometry where this behavior has been found; this gives high importance to this simple geometry for determining the ability of NR compounds to strain-crystallize under a severe stress concentration.

Perhaps, the 20 years it took from the time of automobile invention to the use of carbon black in tires is due to the reduction in tear strength of NR, when it contains low levels of black. One would be hesitant to try higher black levels, if low levels were unsatisfactory. Why does $\boldsymbol{\sigma_{bc}}$ decrease for the lightly filled NR vulcanizate? There must be competing effects.

Gent[25] studied the thermal crystallization of gum and filled NR vulcanizates. He found that filled NR crystallized significantly slower than the gum, and attributed this behavior to a reduction in chain mobility due to carbon black "obstacles." We, thus, propose that the reduced crystallinity and strength with 12 phr of black are due to hindered chain mobility, which inhibits the formation of crystal nuclei. But, why does NR with enough filler become stronger than the gum and at what level of black does knotty tearing commence?

Figure 8 shows $\boldsymbol{\sigma_{bc}}$ for the gum, and for testpieces filled with 12, 14, or 15 phr of black. Only the line for the gum (from Figure 4) is plotted, while all the data for 12, 14, and 15 phr of black are shown. Test pieces with 14 phr of black give results much like those obtained with 12 phr. All fractures occur by growth of a single crack. However, with 15 phr of black, strong and weak populations of $\boldsymbol{\sigma_{bc}}$ are seen when 0.98 mm < c < 2.16 mm. In the weak population, $\boldsymbol{\sigma_{bc}}$ is perhaps slightly greater with 15 phr of black than values for 12 and 14 phr of black; moreover, fracture here occurs by the growth of a single crack across specimens. However, about 50% of the 15 phr specimens in the dual strength range exhibit knotty tear (Figure 8); these are about ***four times*** as strong as the weaker population! Thus, there is fracture instability when the NR contains 15 phr of black. It is remarkable that just 1 phr of black can lead to such a large difference in strength. This sideways cracking ("super-blunting") seems to develop because of sufficient matrix orientation and enchainment of filler particles during deformation. This may give "continuity" to the reinforcing particles and thus block forward crack growth.

The co-existence of weak and strong populations at 15 phr of black is an inherent instability, ***not*** related to differences in composition or processing among specimens. Similarly, a thin column may buckle or not buckle, when placed under a critical compressive load (Eulerian buckling). Rubber also exhibits another ***elastic*** instability. When first inflating a long, small diameter balloon, the diameter increases uniformly. But, with enough inflation an aneurysm forms, and now, the balloon has two diameters at one pressure.

What has occurred at 15 phr of black that has caused knotty tearing and the dramatic increase in tear resistance? Bound rubber tests for samples containing 14 or 15 phr of black were carried-out. With 14 phr, the uncured NR/black composition disintegrates into a suspension after placing it in toluene. On the other hand, an uncured sample with 15 phr of black, swells, but remains a coherent piece which contains all the black and the bound rubber; 60% of the rubber is bound. The rubber that is unbound dissolves into the solvent. Mastication with sufficient volume of very finely divided carbon black joins chains and results in a mass which, when immersed in solvent, separates into a coherent gel (containing solvent, all the carbon black, and [bound] higher molecular weight chains). To form a coherent gel, chain-filler bonding must be solvent resistant and filler particles must be sufficiently close to one another so that: (1) single chains adsorb on adjacent particles, and/or (2) different chains, which are adsorbed on adjacent particles, form entrapped tangles. High reinforcement requires dual network formed by "superimposing" a permanent, elastic chain/chain network onto an impermanent, viscous carbon black/chain network – resulting in a reinforced composition, which is less elastic, but much more resistant to tearing, than the gum network.

A NR/carbon black vulcanizate must possess a percolating black-rubber network in addition to the rubber-rubber chain network to give knotting and outstanding tear strength. When a bound rubber network is present, the hinderance to crystallization at low black levels is overcome, as the bound rubber network provides substantial chain alignment, strain-crystallization, and anisotropy at a cut tip. The result is knotty tearing. The rubber-rubber network is responsible for retraction, while the motion of the black-rubber network contributes energy dissipation.

Another consequence of bound rubber is a rapid increase in stiffness when the amount of filler exceeds the percolation limit. This is demonstrated in Figure 10, where 100% modulus is plotted versus volume fraction of carbon black. There is a rapid upturn in stiffness when the amount of black exceeds 15 phr.

Next, $\boldsymbol{\sigma_{bc}}$ of the NR vulcanizate containing 18 phr of black is shown (Figure 11). The percentage of samples that show knotty tearing increases to about 90% (co-strong population, $SP_{co}$) and the breadth of the region of knotty tearing increases to 0.36 mm < c < 2.15 mm.

Finally, returning to Figure 3, it is noted again that at 50 phr of black all specimens exhibit knotty tear.

We have also carried-out similar studies of NR vulcanizates that contain coarser carbon blacks. When particle size is larger, a larger concentration of black is needed to attain bound rubber. But, as found here, the onset of bound rubber formation and knotty tearing still coincide. Additionally, when the black is sufficiently coarse, the decrease in strength at low black levels does not occur; e.g., at low black levels coarse N990 gives vulcanizates with higher tear strength than ones with N115! This suggests the use of blends of types, when attempting to have low hysteresis, but high tear strength.

G. Rong and J. Jiang conducted the experimental work, L. Jia and G.R. Hamed were responsible for the written text, and N. Schmitz was responsible for the editing and correction.

<u>Additional Information</u>

There are no competing interests.

Table 1. Parts per hundred rubber (phr) of black and volume fractions **v** of black for NR vulcanizates.

| N115 (phr) | 0 | 12 | 14 | 15 | 18 | 50 |
|---|---|---|---|---|---|---|
| **v** | 0 | 0.057 | 0.065 | 0.07 | 0.083 | 0.2 |

Table 2. Internal mixer procedure.

| Step | Time (min) | Component Added |
|---|---|---|
| 1 | 0-1 | Natural Rubber |
| 2 | 1-1.5 | ½ carbon black |
| 3 | 1.5-3 | ZnO, Stearic acid, PVI, DQ, PD-2 and other ½ carbon black |
| 4 | 3-5 | Wax |
| 5 | 5 | Dump and weight |

Table 3. Milling procedure.

| Time (min) | Procedure |
|---|---|
| 0-2 | Mix the two masterbatches together with the nip at 1.9 mm |
| 2-3 | Form rolling bank at 0.6 mm nip |
| 3-6 | Add curing agents (S and TBBS) at 0.6 mm nip |
| - | 10 end roll passes at 1.2 mm nip. Sheet off at a nip setting of about 1.9 mm. Dump. |

Figure Captions

Figure 1. Stress-strain curves for gum and black-filled NR and SBR vulcanizates[9] and for an SBR vulcanizate that contains 50 phr of $CaCO_3$.

Figure 2. Simple model showing spherical inclusions of radius **r**, evenly spaced on a square lattice. Parameters include particle spacing, **s**; diagonal particle separation, **z**; volume fraction of rubber within a distance **t** of the surface of all inclusions.

Figure 3. Schematic showing network chain dimensions relative to particle size, when **d** = 20 nm and **v** = 0.25.

Figure 4. Strength $\boldsymbol{\sigma_{bc}}$ of gum NR and black-filled (50 phr) vulcanizates at various cut depths. Top dotted lines are strengths $\boldsymbol{\sigma_{bo}}$ when no cut is present.

Figure 5. Cracking pattern of specimen #16 in Figure 4. Two broken halves were placed in proximity prior to taking photo. A1-A3 are secondary cracks formed before the catastrophic one, C4.

Figure 6. Mechanism of knotty tearing.

Figure 7. Strength $\boldsymbol{\sigma_{bc}}$ of gum NR and black-filled (12 phr) vulcanizates at various cut depths. Top dotted lines are strength $\boldsymbol{\sigma_{bo}}$ when no cut is present.

Figure 8. Strength $\sigma_{bc}$ of gum NR and black-filled (12 phr, 14 phr, 15 phr) vulcanizates at various cut depths. Top dotted lines are strengths $\boldsymbol{\sigma_{bo}}$ when no cut is present.

Figure 9. Cracking pattern of specimen #18 in Figure 8. Two broken halves were placed in proximity prior to taking photo. A1 and A2 are secondary cracks formed before the catastrophic one C3.

Figure 10. Modulus at 100% for various volume percent of black. 15 phr of black is denoted.

Figure 11. Strength of gum NR and a black-filled (18 phr) vulcanizate at various cut depths. Top dotted lines are strengths $\sigma_{bo}$ when no cut is present.

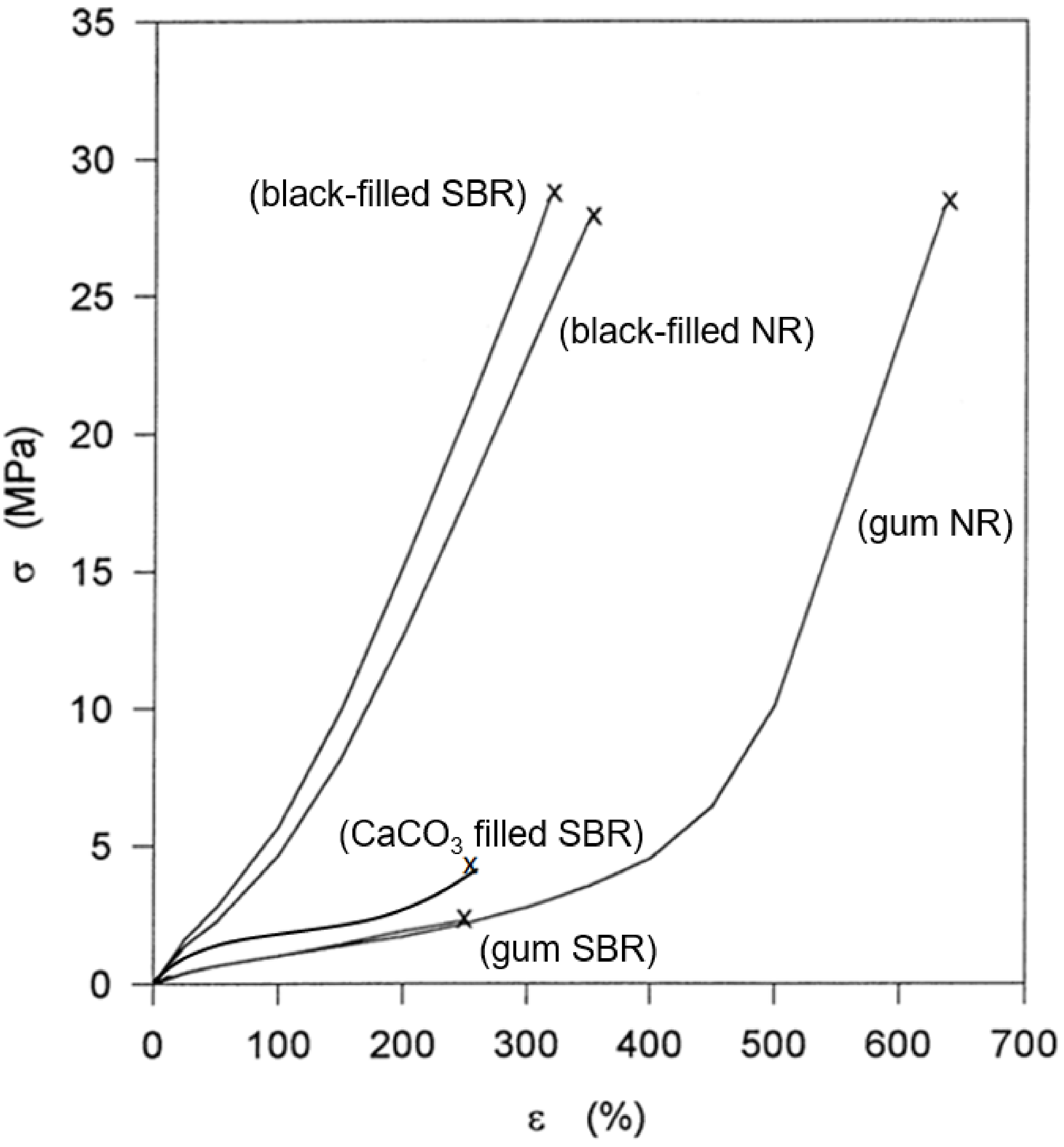

Figure 1.

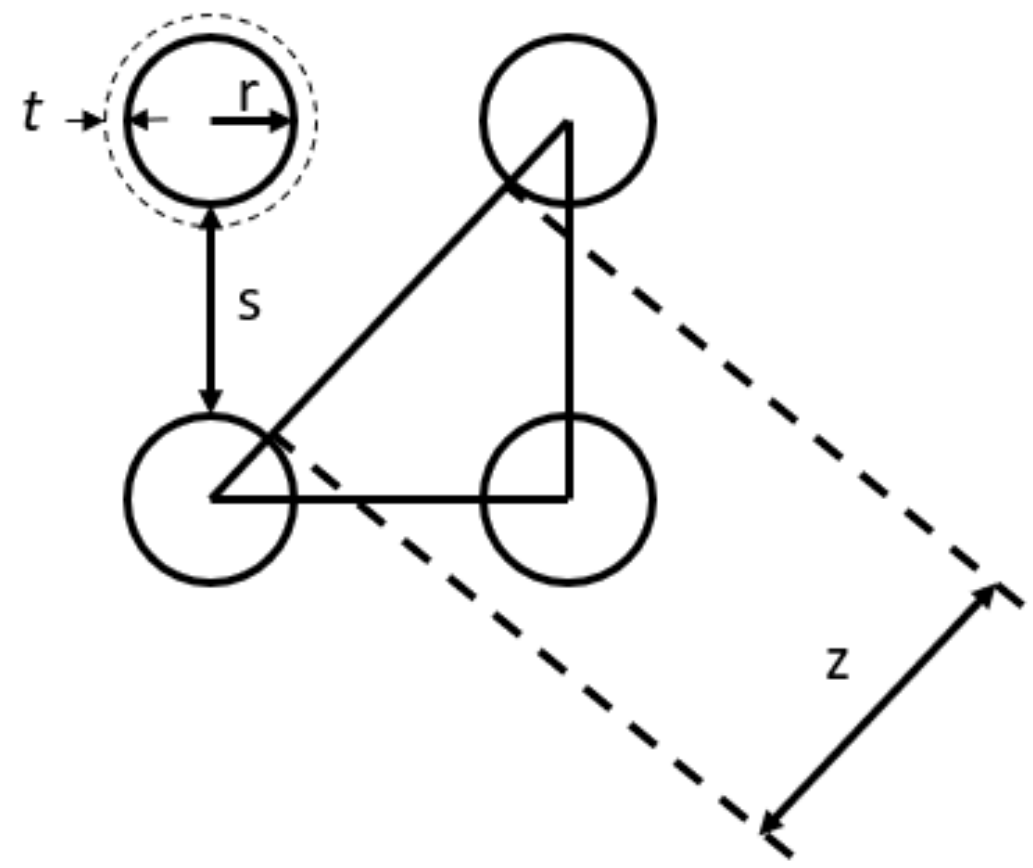

$v$ : volume fraction of filler
$v_t$ : volume fraction of rubber within a distance t of the particles
r: the radius of filler
s: the spacing between nearest filler
z: the diagonal spacing between nearest filler

- particles placed on a simple square cubic lattice

$$s = r \times \left[ \frac{1.612}{v^{1/3}} - 2 \right]$$

$$z = r \times \left[ \frac{2.28}{v^{1/3}} - 2 \right]$$

$$v_t = \frac{v \left[ \left( 1 + \frac{t}{r} \right)^3 - 1 \right]}{1 - v}$$

Figure 2.

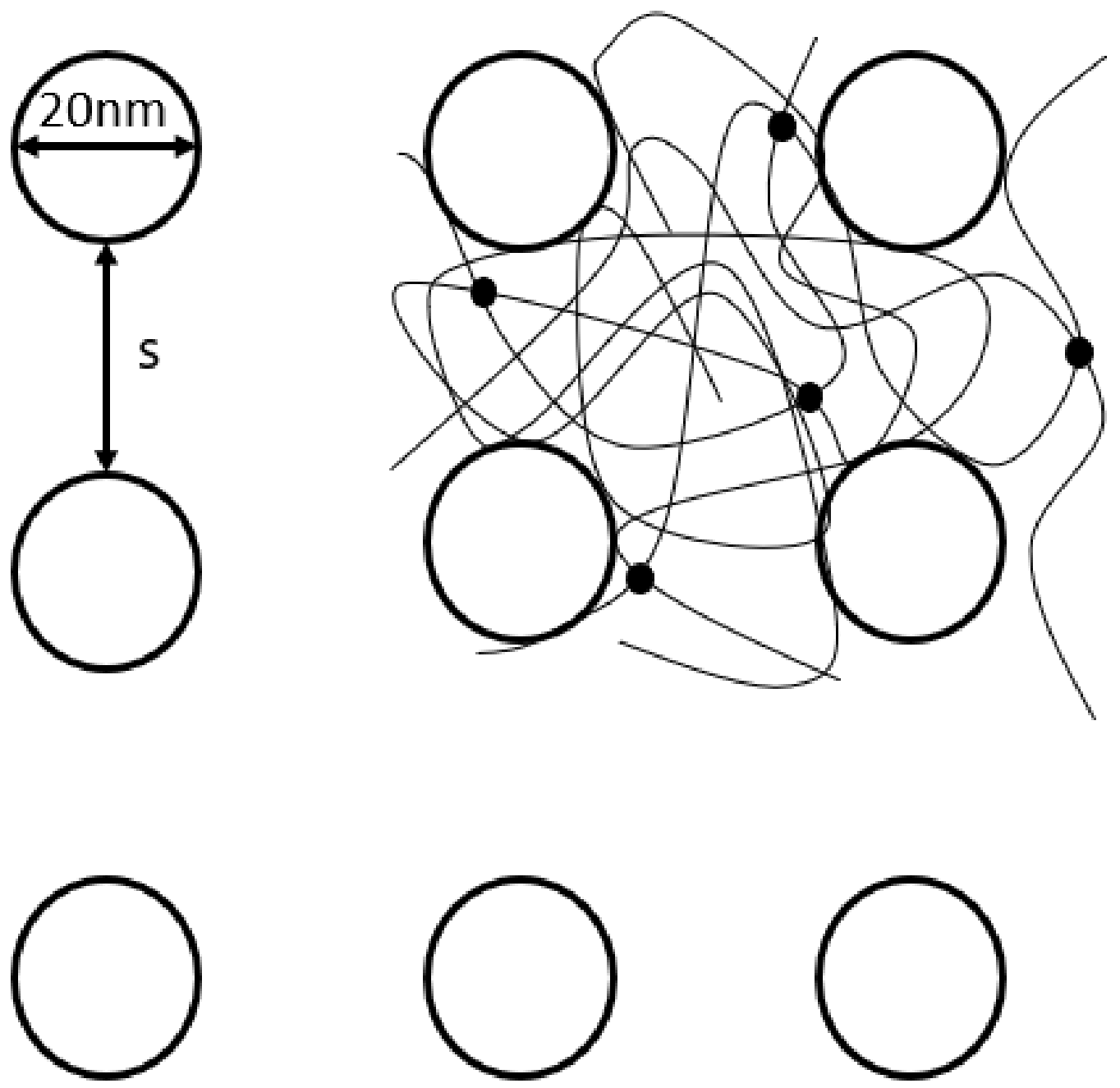

Figure 3.

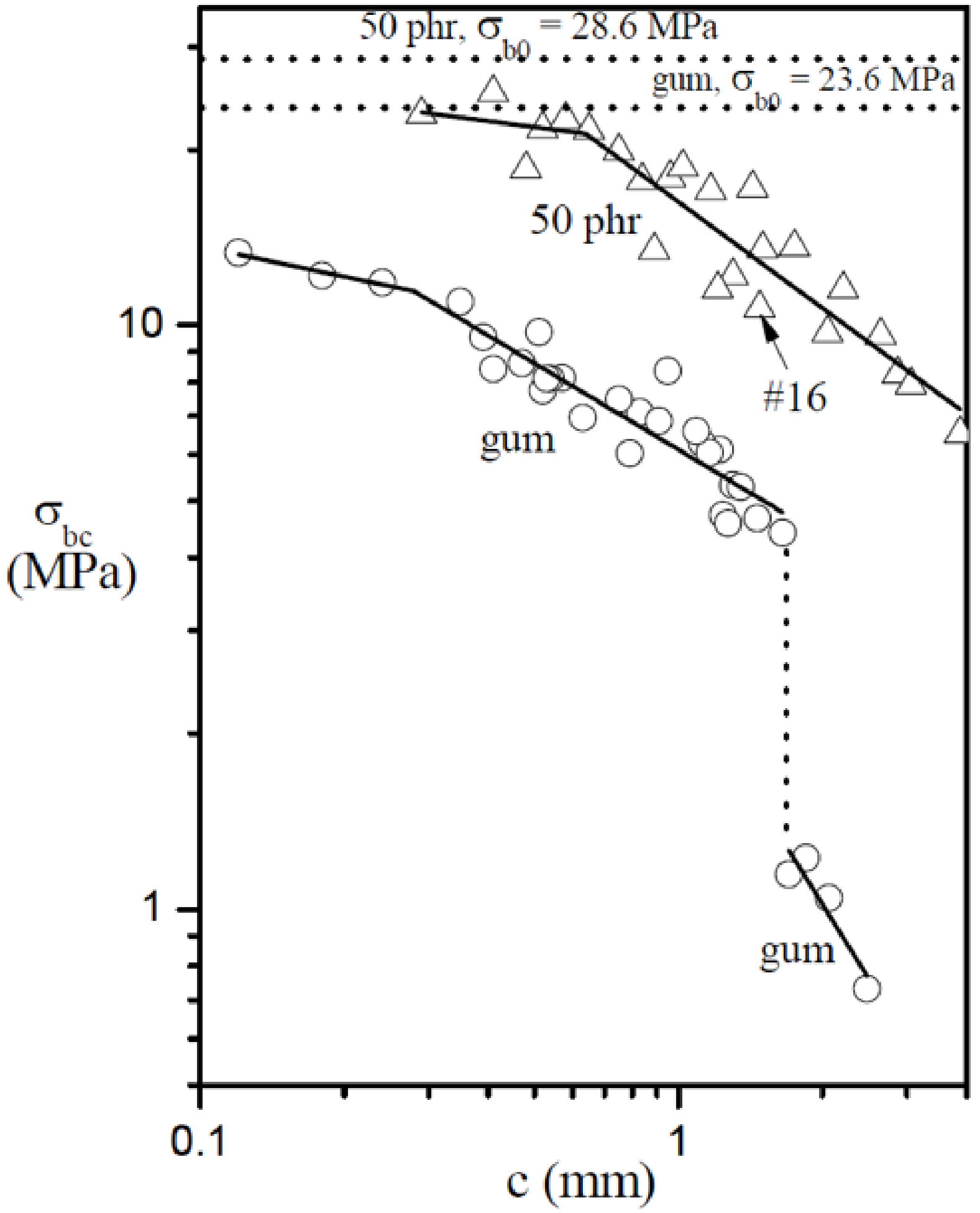

Figure 4.

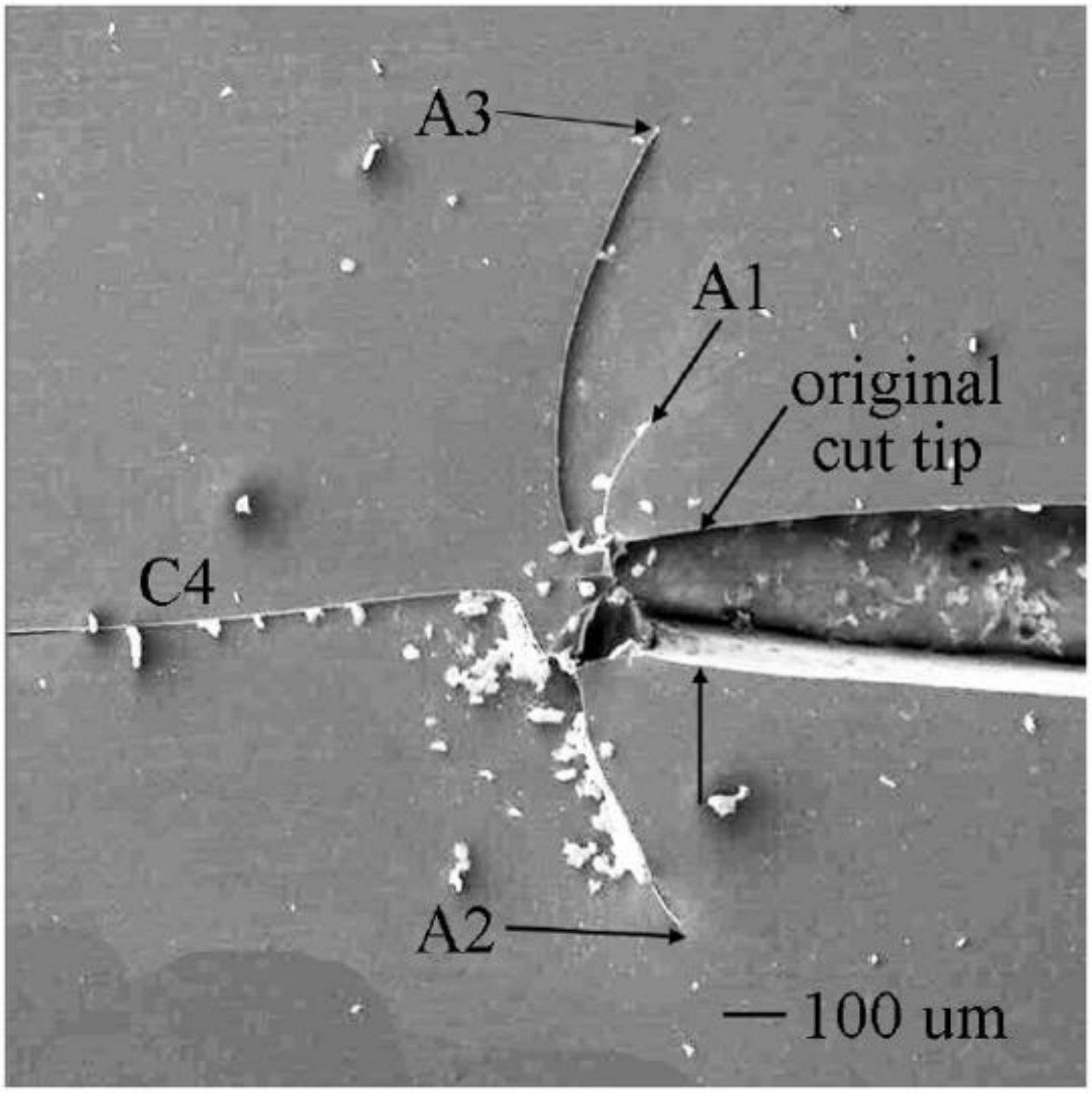

Figure 5.

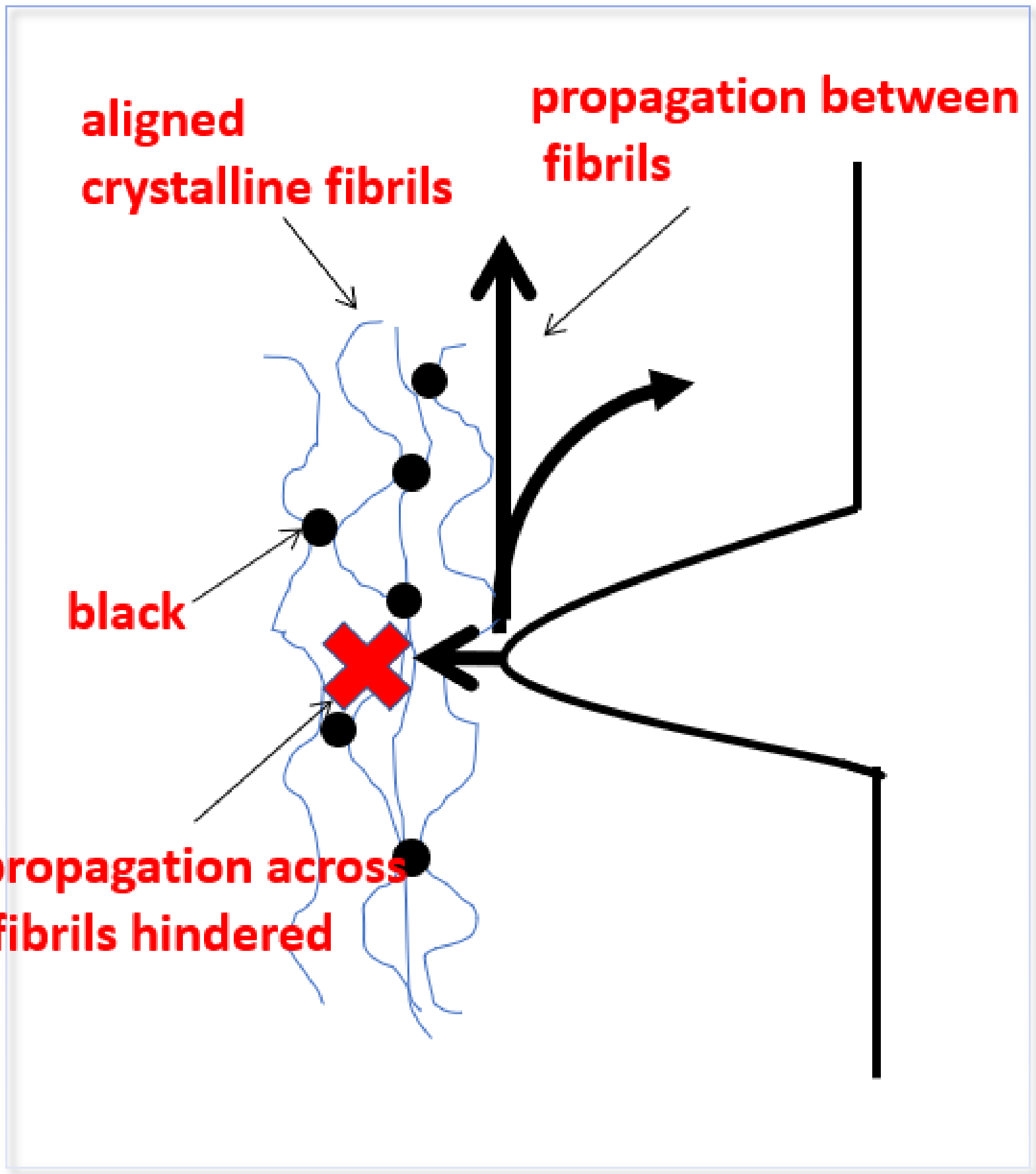

Figure 6.

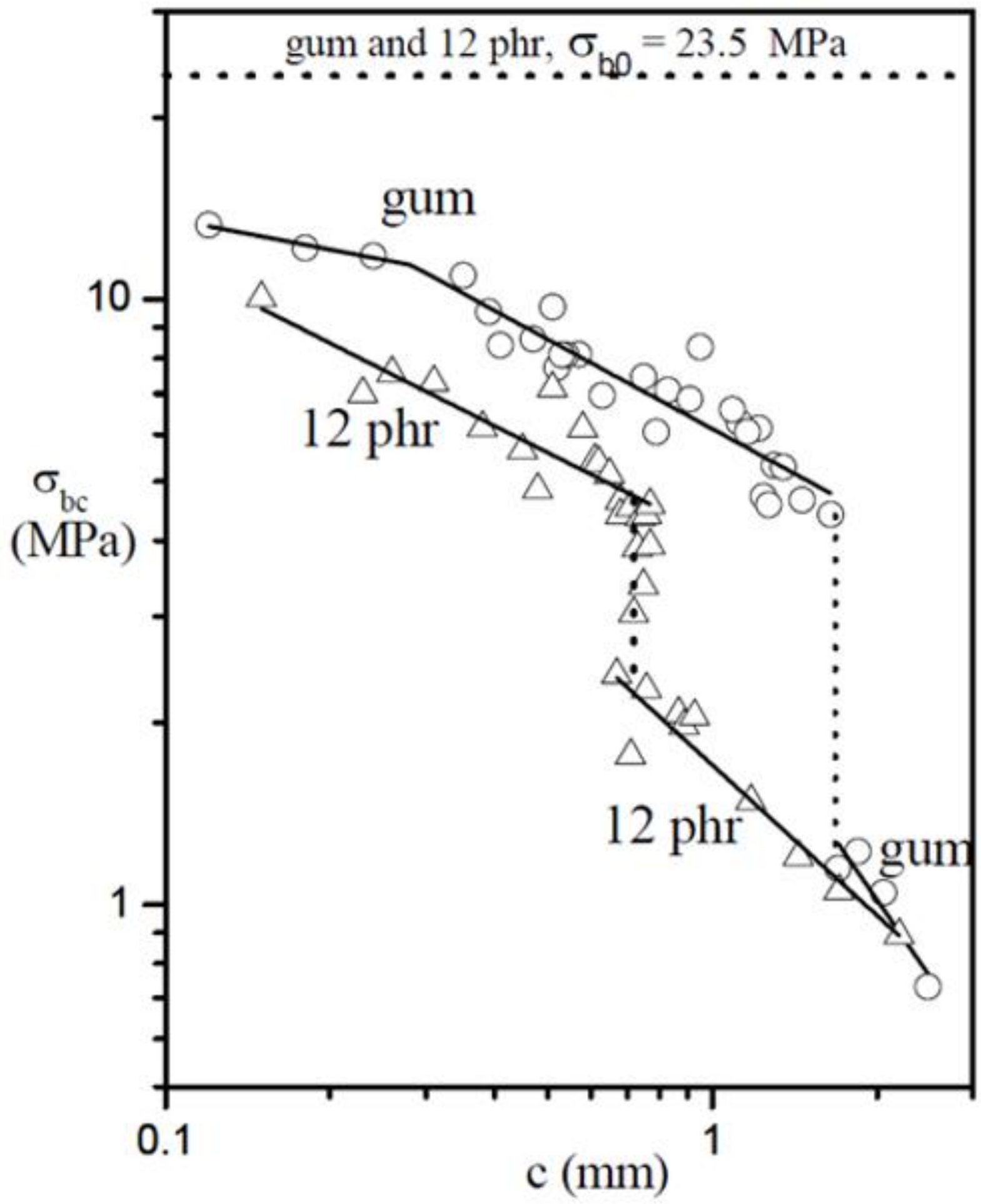
gum and 12 phr, $\sigma_{b0}$ = 23.5 MPa
gum
12 phr
12 phr
gum
$\sigma_{bc}$ (MPa)
10
1
0.1
1
c (mm)

Figure 7.

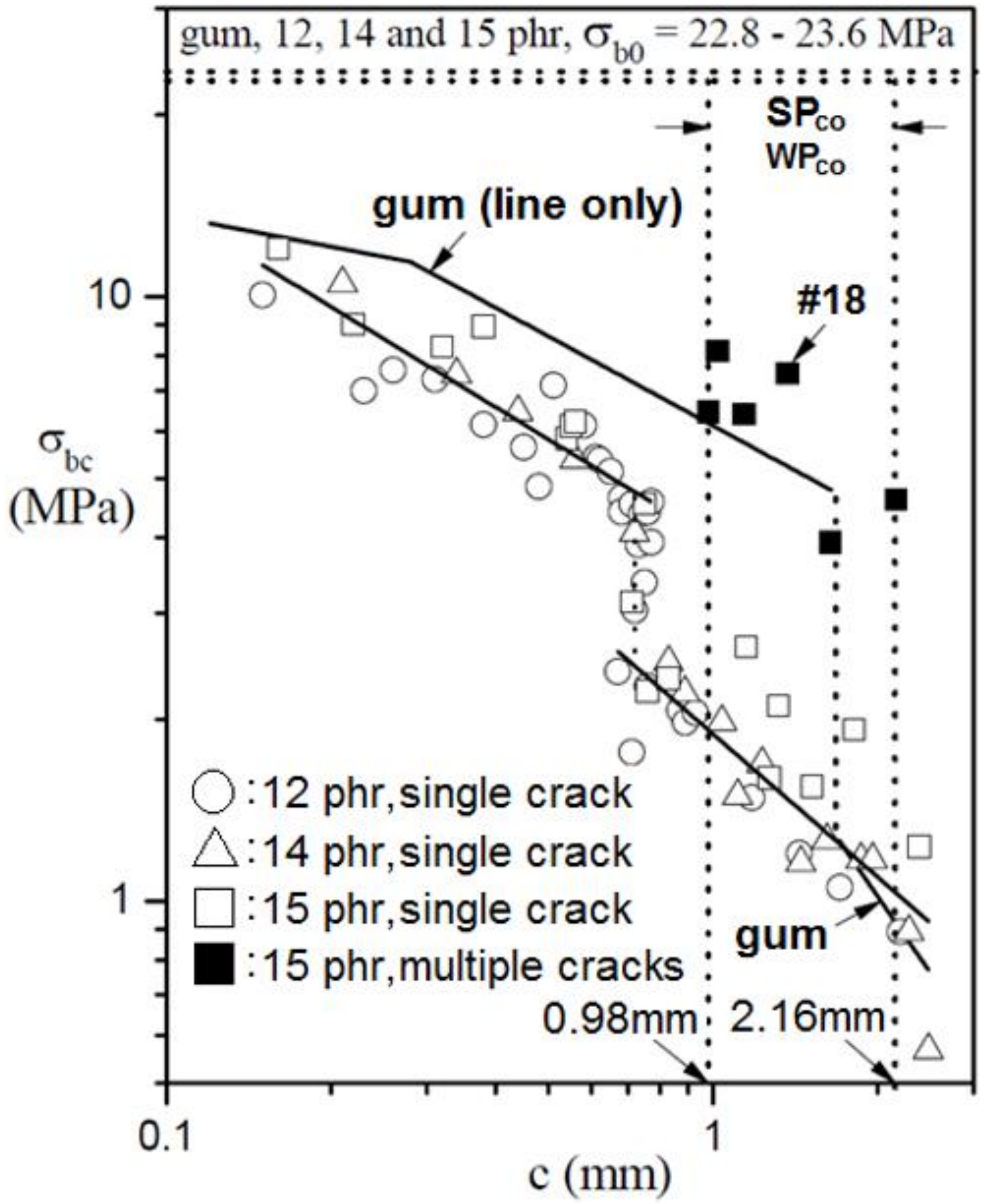
gum, 12, 14 and 15 phr, $\sigma_{b0}$ = 22.8 - 23.6 MPa
SPco
WPco
gum (line only)
#18
$\sigma_{bc}$ (MPa)
10
1
:12 phr,single crack
:14 phr,single crack
:15 phr,single crack
:15 phr,multiple cracks
gum
0.98mm
2.16mm
0.1
1
c (mm)

Figure 8.

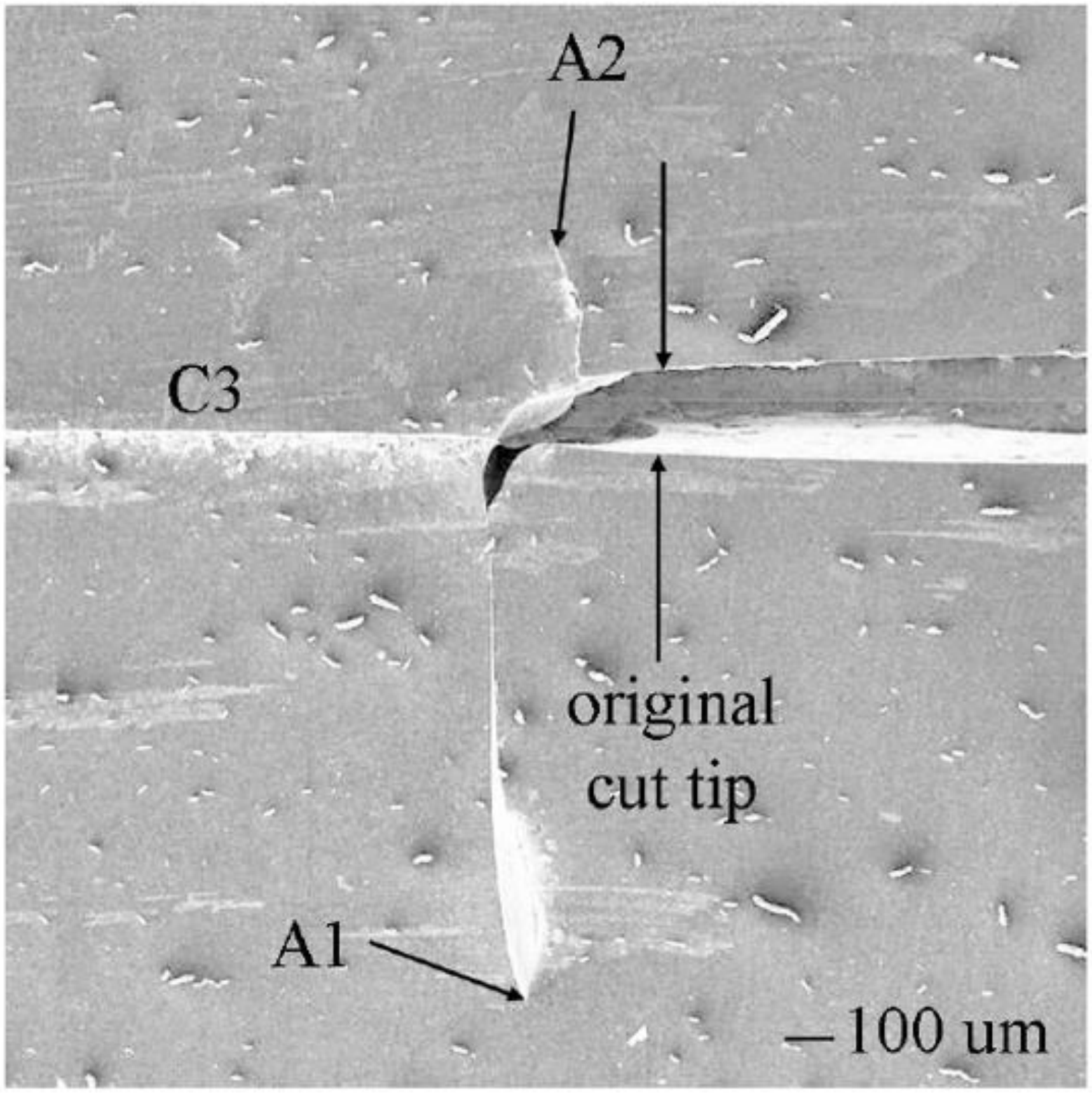

Figure 9.

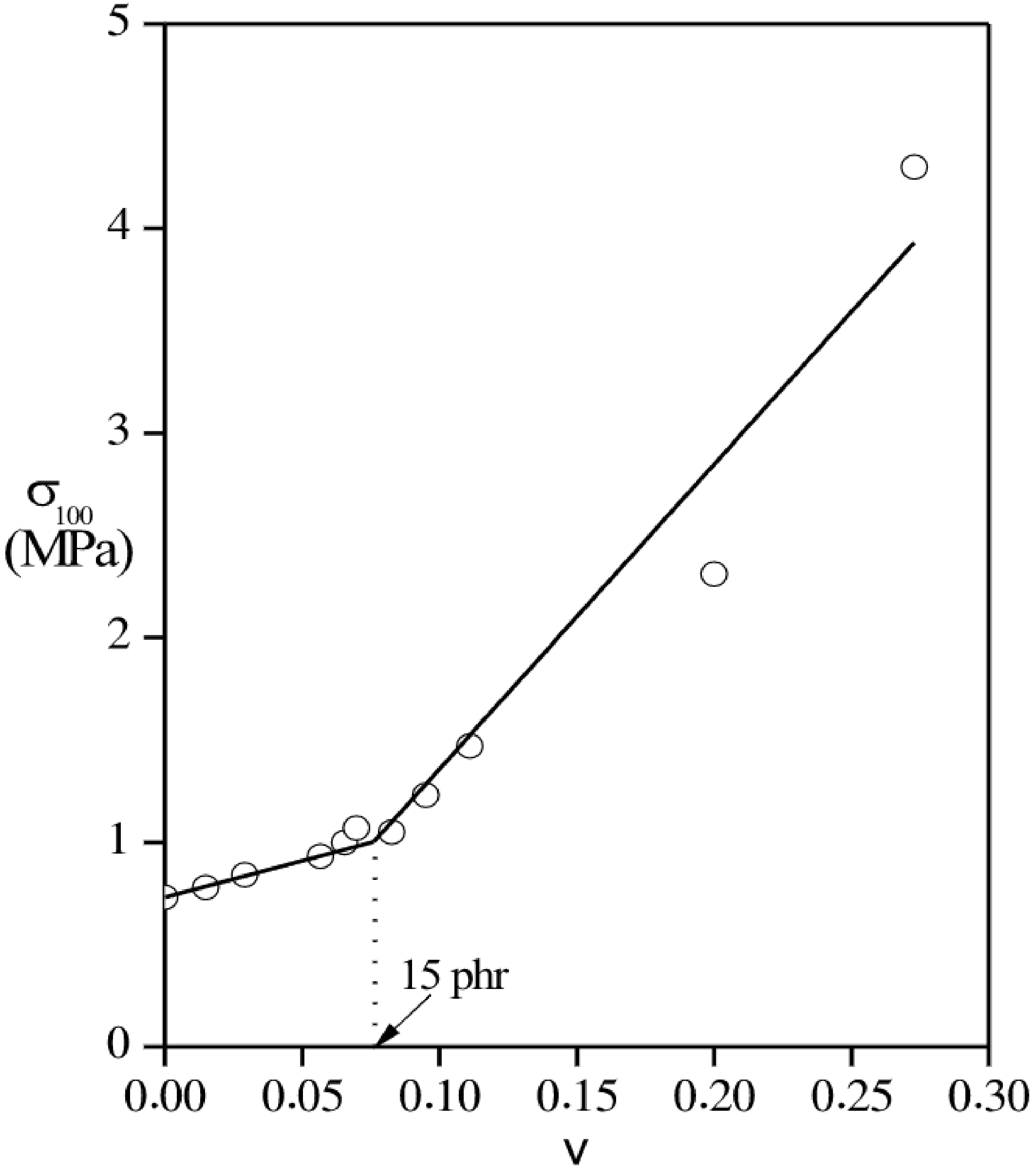

Figure 10.

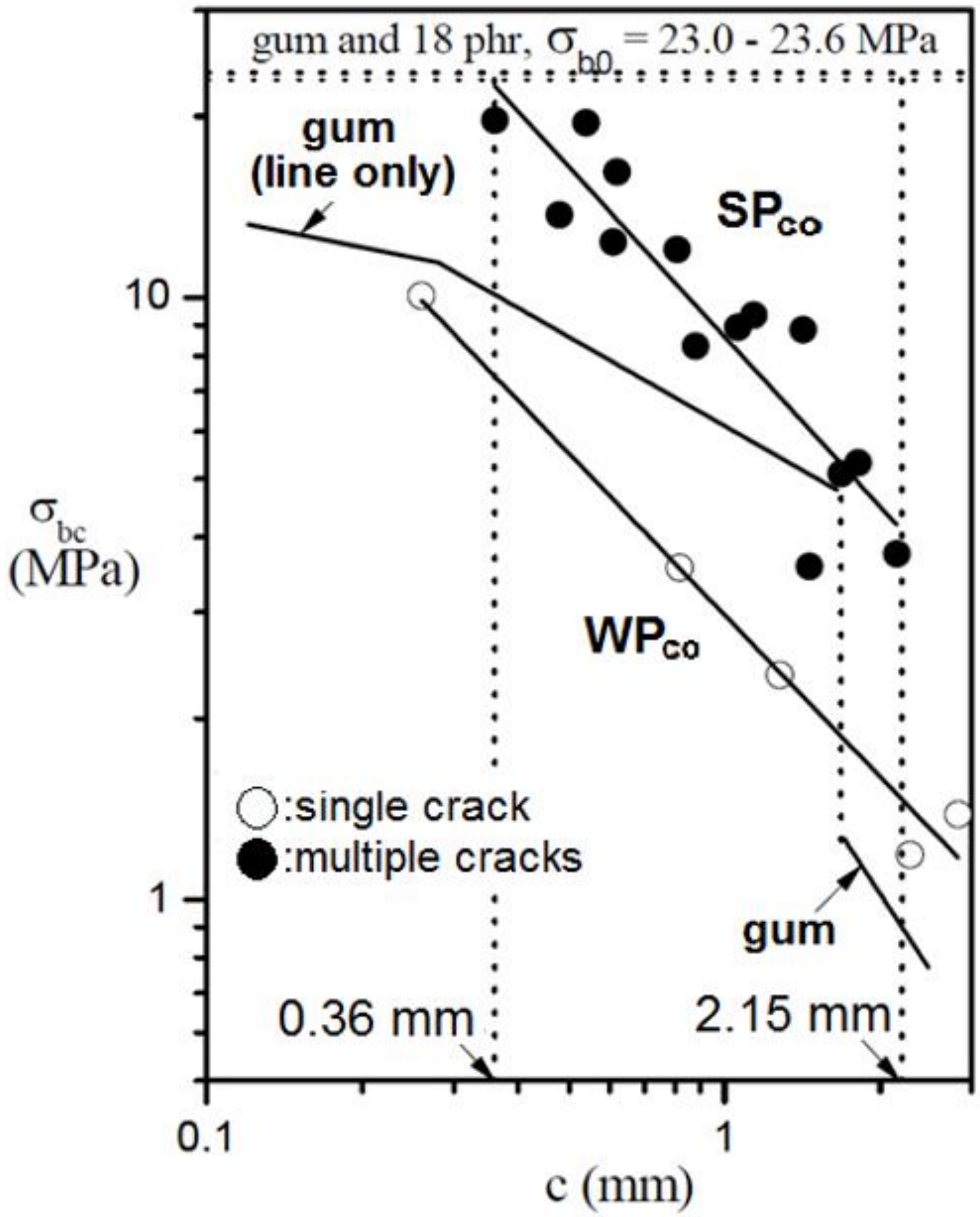
gum and 18 phr, σb0 = 23.0 - 23.6 MPa
gum
(line only)
SPco
WPco
10
1
σbc
(MPa)
○:single crack
●:multiple cracks
gum
0.36 mm
2.15 mm
0.1
1
c (mm)

Figure 11.